\documentclass[prd,twocolumn,superscriptaddress]{revtex4}
\usepackage{amsmath}
\usepackage{adjustbox}
\usepackage{gensymb}
\usepackage{amsfonts}
\usepackage{CJKutf8}
\usepackage{graphicx}
\usepackage{hyperref}
\usepackage{amssymb}
\usepackage{cancel}
\usepackage{color}
\usepackage{amscd}
\usepackage{bm}
\usepackage{adjustbox}
\makeatletter
\renewcommand{\maketag@@@}[1]{\hbox{\m@th\normalsize\normalfont#1}}%
\makeatletter
\begin{document}
\begin{CJK*}{UTF8}{gbsn}
\title{Validity of the Background Subtraction Method for Black Hole Thermodynamics in Matter–Coupled Gravity Theories}
\author{Yong Xiao}
\email{xiaoyong@hbu.edu.cn}
\affiliation{Key Laboratory of High-precision Computation and Application of Quantum Field Theory of Hebei Province,
College of Physical Science and Technology, Hebei University, Baoding 071002, China}
\affiliation{Hebei Research Center of the Basic Discipline for Computational Physics, Baoding, 071002, China}
\affiliation{Higgs Centre for Theoretical Physics, School of Mathematics, University of Edinburgh, Edinburgh, EH9 3FD, United Kingdom}
\author{Aonan Zhang}
\affiliation{Key Laboratory of High-precision Computation and Application of Quantum Field Theory of Hebei Province,
College of Physical Science and Technology, Hebei University, Baoding 071002, China}
\affiliation{Hebei Research Center of the Basic Discipline for Computational Physics, Baoding, 071002, China}

\begin{abstract}
The background subtraction method has long served as a practical tool for computing the Euclidean action and thermodynamic quantities of black holes. While its equivalence to the Iyer--Wald formalism is well understood in pure gravity theories, its validity in matter-coupled theories remains less clear and has even been questioned in the literature. In this work, we revisit this issue and demonstrate that the equivalence between the Euclidean action method and the Iyer--Wald formalism persists in matter-coupled scenarios. We apply the resulting formulation to two representative examples of such theories, and in both cases, the Euclidean approach performs smoothly. We further identify situations where the method may encounter subtleties due to the special properties of certain matter fields. Our results clarify when background subtraction remains reliable beyond pure gravity and when additional care is necessary.
\end{abstract}

 \maketitle

\section{Introduction}

To analyze black hole thermodynamics, a key step often involves computing the Euclidean action of a given black hole. Consider a Lagrangian of the form
\begin{align}
    \mathbf{L} = L\,\boldsymbol{\epsilon},
\end{align}
where $\boldsymbol{\epsilon}$ denotes the volume form of a $D=d+1$ dimensional spacetime (we use boldface to represent differential forms throughout this work). For black hole solutions in asymptotically anti-de Sitter (AdS) spacetimes, direct evaluation of the bulk integral $\int_{\mathcal{M}}^{(\mathrm{BH})} \boldsymbol{L}$, together with its associated boundary term $\int_{\partial\mathcal{M}}^{(\mathrm{BH})} \boldsymbol{B}$, yields a divergent result. Consequently, an appropriate regularization procedure is required to remove such divergences.

Following the foundational works~\cite{Gibbons:1976ue,hawkingpage},  the Euclidean action of AdS black holes can be regularized by subtracting the contribution of a reference background. This background is typically chosen to be pure AdS spacetime with a properly redshifted Euclidean time. Over the past decades, the resulting background subtraction method has been widely applied to extract thermodynamic quantities for a broad class of black holes \cite{Gubser:1998nz,Gibbons:2004ai,Dutta:2006vs,Xiao:2023two}. In this approach, the Euclidean action is defined as
\begin{equation}
    I_E \equiv 
    \int_{\mathcal{M}}^{(\mathrm{Reg.})} \boldsymbol{L}_E
    +
    \int_{\partial\mathcal{M}}^{(\mathrm{Reg.})} \boldsymbol{B}_E ,
    \label{eucl_formula}
\end{equation}
where the notation $\int^{(\mathrm{Reg.})} \equiv \int^{(\mathrm{BH})} - \int^{(\mathrm{bg})}$ abbreviates the regularized integral, with superscripts ``$BH$" and  ``$bg$"  respectively denoting the black hole spacetime and reference background. Here, $\boldsymbol{L}_E$ represents the Euclideanized Lagrangian, and $\boldsymbol{B}_E$ denotes the corresponding Euclidean boundary term.

For pure gravity theories in AdS spacetime, the boundary term contribution always vanishes after background subtraction, so Eq.~\eqref{eucl_formula} reduces to
\begin{equation}
    I_E =
    \int_{\mathcal{M}}^{(\mathrm{Reg.})} \boldsymbol{L}_E .
    \label{eucl2}
\end{equation}
This shows that the background subtraction method leads to a very simple computation: one merely inserts the black hole metric into the Lagrangian and performs the integration. Compared with the covariant counterterm method (or holographic renormalization)~\cite{Balasubramanian:1999re,Bianchi:2001kw,Papadimitriou:2005ii}, the background subtraction method avoids constructing counterterms, which is often highly nontrivial for higher-derivative gravity theories.

On the other hand, the Iyer–Wald formalism is another powerful method for studying black hole thermodynamics~\cite{Wald:1993nt,Iyer:1994ys,Iyer:1995kg}. These early works have already explained that the results obtained from the Euclidean method are consistent with the Noether–charge construction. Although the essential idea was known long ago, a refined and fully unambiguous explanation of how background subtraction emerges naturally from the Iyer–Wald formalism has appeared only recently in~\cite{Guo:2025ohn,Guo:2025muo,Chen:2025ary}.

In Sec.~\ref{sec20} of this paper, we will give a concise review of the argument presented in~\cite{Guo:2025ohn}, showing explicitly how the background subtraction formulas follow from the Iyer–Wald formalism. While Ref.~\cite{Guo:2025ohn} focused on pure gravity, the same reasoning can be generalized to gravity theories coupled to matter fields. Therefore, in Sec.~\ref{sec:III}, we apply the subtraction method to analyze black hole thermodynamics in matter–coupled theories. Because the derivation is strict and unambiguous, it can be regarded as a diagnostic tool: if the derivation is applicable, it produces the correct thermodynamic relations; if not, it also indicates precisely where the difficulty arises. 

Notably, a controversy exists in the literature: for gravity theories with matter fields non–minimally coupled to curvature, it has been argued that the Euclidean action method may fail to yield correct black hole thermodynamics~\cite{Feng:2015sbw,Cai:2011uh}. With a systematic diagnostic framework now available, we are strongly motivated to revisit this example and examine other representative matter-coupled scenarios. Concluding remarks are presented in Sec.~\ref{sec:conclusion}.

\section{The derivation of background subtraction method from Iyer--Wald formalism} \label{sec20}

Based on the excellent earlier developments~\cite{Wald:1993nt,Iyer:1994ys,Iyer:1995kg,Guo:2025ohn,Guo:2025muo,Chen:2025ary}, we now have a clear understanding of how the background subtraction method can be naturally induced from the Iyer--Wald formalism, and why it is expected to be valid.

We start from a general diffeomorphism-invariant Lagrangian
\begin{equation}
    \mathbf{L}=L\,\boldsymbol{\epsilon},
\end{equation}
The variation of this Lagrangian with respect to the dynamical fields $\phi \equiv \{g_{\mu\nu},\psi\}$ (encompassing the metric and matter fields) takes the form
\begin{equation}
    \delta\mathbf{L} = \mathbf{E}^{\phi}\,\delta\phi + d\boldsymbol{\Theta}[\delta\phi],
    \label{gelvariform}
\end{equation}
where $\mathbf{E}^{\phi} = 0$ are the equations of motion, and $d\boldsymbol{\Theta}$ is a total derivative term. For a fixed vector field $\xi$, the standard Noether current is defined as
\begin{equation}
    \mathbf{J}_\xi \equiv \boldsymbol{\Theta}[\mathcal{L}_\xi\phi] - \xi\cdot\mathbf{L},
    \label{gelj}
\end{equation}
which satisfies $d\mathbf{J}_\xi = 0$ on-shell. This allows us to construct the Noether charge $\mathbf{Q}_\xi$ from
\begin{equation}
    \mathbf{J}_\xi = d\mathbf{Q}_\xi.
\end{equation}
When $\xi$ is taken to be a Killing vector, the Iyer--Wald formalism shows that the following two identities hold:
\begin{align}
    & d\big( \delta\mathbf{Q}_\xi - \xi\cdot\boldsymbol{\Theta}[\delta\phi] \big) = 0,
    \label{ff1}\\
 &   d\mathbf{Q}_\xi = -\,\xi\cdot\mathbf{L}.
    \label{ff2}
\end{align}
To derive these identities, one uses both $\mathbf{E}^\phi=0$ and $\delta\mathbf{E}^\phi=0$. This means that the variation in Eq.~\eqref{ff1} should be understood as a variation within the solution space, usually parameterized by quantities such as the mass $M$, angular momentum $J$, or conserved charge $Q$, etc.

Next we concentrate on the Einstein gravity described by the Lagrangian
\begin{equation}
    \boldsymbol{L} = \frac{1}{16\pi}(R - 2\Lambda)\boldsymbol{\epsilon},
\end{equation}
where $R$ is the Ricci scalar and $\Lambda$ is the negative cosmological constant. For this theory, the Noether charge $\boldsymbol{Q}$ and the $\boldsymbol{\Theta}$ term take the explicit forms \footnote{Our convention for the differential form is $\boldsymbol{\epsilon}_{\mu_1 \cdots \mu_p} \equiv \frac{\sqrt{|g|}}{p! (D-p)!}\epsilon_{\mu_1 \cdots \mu_p\nu_{p+1}\cdots \nu_{D}}dx^{\nu_{p+1}} \wedge\cdots dx^{\nu_{D}}$.}:
\begin{align}
    \boldsymbol{Q}_\xi &= -\frac{1}{16\pi} \left( \nabla^\mu\xi^\nu - \nabla^\nu\xi^\mu \right) \boldsymbol{\epsilon}_{\mu\nu\cdots},\\[1mm]
    \boldsymbol{\Theta}[\delta g] &= \frac{1}{16\pi} \left( g^{\mu\alpha}\nabla^\nu\delta g_{\alpha\nu} - g^{\alpha\beta}\nabla^\mu\delta g_{\alpha\beta} \right) \boldsymbol{\epsilon}_{\mu\cdots}.
\end{align}
We analyze the thermodynamic properties of a four dimensional Schwarzschild–AdS black hole, with the metric
\begin{equation}
    ds^2
    =
    \!-\!\Big(1\!-\!\frac{2m}{r}\!-\!\frac{\Lambda r^2}{3}\Big)dt^2
    +\frac{dr^2}{1\!-\!\frac{2m}{r}\!-\!\frac{\Lambda r^2}{3}}
    + r^2 d\Omega_2.
\end{equation}
This simple case clearly demonstrates the core logic of the analysis without unnecessary complications. The method generalizes straightforwardly to stationary black holes with angular momentum or higher-derivative diffeomorphism-invariant gravity theories.

Taking $\xi_t=\frac{\partial}{\partial t}$ as the timelike Killing vector, Eq.~\eqref{ff1} becomes
\begin{equation}
    d\boldsymbol{k}_{\xi_t} = 0,
\end{equation}
where we define the Iyer--Wald surface charge density
\begin{equation}
    \boldsymbol{k}_{\xi_t}
    \equiv
    \delta\mathbf{Q}_{\xi_t}
    -
    \xi_t\cdot\boldsymbol{\Theta}[\delta\phi]. \label{sec2wdef}
\end{equation}
Integrating $d\boldsymbol{k}_{\xi_t}$ over a spatial hypersurface $\Sigma$ that extends from the bifurcation surface $S_{r_h}$ to a large-$r$ surface $S_\infty$, Gauss's law yields
\begin{equation}
    \int_{S_{r_h}}\boldsymbol{k}_{\xi_t} = \int_{S_\infty}\boldsymbol{k}_{\xi_t}.
    \label{id1}
\end{equation}
In practical calculations, an infrared cutoff $r_c$ should be introduced to regulate divergences. Quantities are computed at $r = r_c$, and the limit $r_c \rightarrow \infty$ is taken after canceling divergent behaviors.

For the static case where $\xi_t$ is the horizon Killing vector, the left-hand side of Eq.~\eqref{id1} is known to equal $T\,\delta S$, where $T$ is the Hawking temperature and $S$ is the black hole entropy \cite{Wald:1993nt,Iyer:1994ys}. For the right-hand side, we can evaluate $\boldsymbol{\Theta}$ on the timelike boundary $\Gamma$ at infinity, where $S_\infty = \Sigma \cap \Gamma$ and the induced volume form $\hat{\boldsymbol{\epsilon}}$ on $\Gamma$ satisfies $\boldsymbol{\epsilon} = \boldsymbol{n}\wedge \hat{\boldsymbol{\epsilon}}$ (where $\boldsymbol{n}\propto \nabla r$ being the outward unit normal vector to $\Gamma$). Following standard procedures \cite{Padmanabhan:2014lwa,Chakraborty:2017zep,Jiang:2018sqj}, $\boldsymbol{\Theta}$ at the boundary can be decomposed as
\begin{equation}
    \boldsymbol{\Theta}[\delta\phi]\big|_{\Gamma} = -\delta\boldsymbol{B} + d\boldsymbol{C}[\delta\phi] + \boldsymbol{F}[\delta\phi]. \label{sec2theta1}
\end{equation}

For Einstein gravity, this decomposition becomes explicitly
\begin{equation}
    \boldsymbol{\Theta}\big|_{\Gamma} = \frac{1}{16\pi}\left( -\delta(2K\hat{\boldsymbol{\epsilon}}) + (K_{\mu\nu} - Kh_{\mu\nu})\delta h^{\mu\nu}\hat{\boldsymbol{\epsilon}} \right),
\end{equation}
allowing us to identify
\begin{equation}
    \boldsymbol{B} = \frac{1}{8\pi}K\hat{\boldsymbol{\epsilon}},
    \qquad
    \boldsymbol{F}[\delta\phi] = \frac{1}{16\pi} (K_{\mu\nu} - Kh_{\mu\nu}) \delta h^{\mu\nu}\hat{\boldsymbol{\epsilon}}.
\end{equation}
Here, $K$ is the extrinsic curvature of $\Gamma$, and $h_{\mu\nu}$ is the induced metric. The $d\boldsymbol{C}$ term also has a well-defined expression but vanishes in calculations, so we omit it throughout this work \footnote{While the $d\boldsymbol{C}$ term vanishes numerically, it is useful for formal derivations, e.g., rewriting Eq.~\eqref{sec2mass} in the Brown-York mass form. See \cite{Harlow:2019yfa,Guo:2024oey} for details.}.

Substituting $\int_{S_h}\boldsymbol{k}_{\xi_t} = T\delta S$ and Eqs.~\eqref{sec2wdef}, \eqref{sec2theta1} into Eq.~\eqref{id1}, we obtain
\begin{equation}
    T\,\delta S = \int_{S_\infty}^{(\mathrm{BH})} \delta\left( \boldsymbol{Q}_{\xi_t} + \xi_t\cdot\boldsymbol{B} \right) - \xi_t\cdot\boldsymbol{F}(\delta\phi).
    \label{iwadsbh}
\end{equation}
Background subtraction provides a systematic way to get rid of the annoying $\int \xi_t\cdot \boldsymbol{F}$ term. We introduce a redshifted pure AdS background as the reference spacetime, with the metric
\begin{equation}
    ds^2 = -\left(1 - \frac{\Lambda r^2}{3}\right)\lambda(r_c)^2 dt^2 + \frac{dr^2}{1 - \frac{\Lambda r^2}{3}} + r^2 d\Omega_2.
\end{equation}
 At $r=r_c \rightarrow \infty$, the redshift factor $\lambda(r_c)$ should be chosen as the form 
\begin{equation}
    \lambda(r_c) = \sqrt{ \frac{g^{(\mathrm{BH})}_{tt}(r_c)}{g^{(AdS)}_{tt}(r_c)} } = \sqrt{ \frac{1 - \frac{2m}{r_c} - \frac{\Lambda r_c^2}{3}}{1 - \frac{\Lambda r_c^2}{3}} }. \label{redlam}
\end{equation}
The redshift factor \eqref{redlam} is widely used in the literature for evaluating the Euclidean action \cite{Gibbons:1976ue,hawkingpage,Gubser:1998nz,Gibbons:2004ai,Dutta:2006vs,Xiao:2023two}, with the key physical motivation for this choice being that it ensures the black hole spacetime and pure AdS spacetime have identical asymptotic geometries (i.e., identical induced metrics) in the limit $r=r_c\rightarrow\infty$. Crucially, it enables us to establish a bridge between the conventional Iyer--Wald formalism and the Euclidean background subtraction method \footnote{An alternative background metric choice with $\lambda(r_c)=1$ exists; though useful for analyzing the Smarr relation, it is irrelevant to the present discussion.}. For this horizonless background spacetime, we have $0=\int_{S_\infty}\boldsymbol{k}_{\xi_t} $, leading to
\begin{equation}
    0 = \int_{S_\infty}^{(\mathrm{bg})} \delta\left( \boldsymbol{Q}_{\xi_t} + \xi_t\cdot\boldsymbol{B} \right) - \xi_t\cdot\boldsymbol{F}(\delta\phi).
    \label{iwads0}
\end{equation}
It can be verified that the $\boldsymbol{F}$ terms cancel between the black hole spacetime and the background at $r_c \rightarrow \infty$:
\begin{equation}
    \int_{S_{r_c}}^{(\mathrm{BH})}\xi_t\cdot\boldsymbol{F}(\delta\phi) - \int_{S_{r_c}}^{(\mathrm{bg})}\xi_t\cdot\boldsymbol{F}(\delta\phi) \rightarrow 0.
\end{equation}

Accordingly, subtracting Eq.~\eqref{iwads0} from Eq.~\eqref{iwadsbh}, we find
\begin{equation}
    T\,\delta S = \delta \int_{S_\infty}^{(\mathrm{Reg.})} \left( \boldsymbol{Q}_{\xi_t} + \xi_t\cdot\boldsymbol{B} \right).
    \label{sec2firstlaw}
\end{equation}
Comparing it with the first law of black hole thermodynamics $T\,\delta S = \delta M$, we identify the mass as
\begin{equation}
    M \equiv \int_{S_\infty}^{(\mathrm{Reg.})} \left( \boldsymbol{Q}_{\xi_t} + \xi_t\cdot\boldsymbol{B} \right).
    \label{sec2mass}
\end{equation}
For the Schwarzschild-AdS black hole example, one can insert the expressions of $\boldsymbol{Q}_{\xi_t}$ and $\boldsymbol{B}$ into Eq.~\eqref{sec2mass} and verify that it indeed gives the expected mass $M=m$. This expression can also be rewritten in the ADM or Brown–York form, but that lies beyond our present scope; see~\cite{Iyer:1994ys,Harlow:2019yfa,Guo:2024oey} for details. Besides, the term $T\delta S$ in Eq.\eqref{sec2firstlaw} appears to be non-integrable at first glance; however, it is indeed integrable, as the simplest Schwarzschild-AdS spacetime depends on only a single parameter. This fact becomes transparent when we express it in the form $T(M) \delta S(M)$.

Next, we demonstrate how the quantum statistical relation $I_E = \beta (M-TS_W) $ arises naturally, where $\beta\equiv 1/T$ is the inverse of Hawking temperature and the periodicity of the Euclidean time $\tau$. The derivation is as follows:
\begin{align}
\begin{split}
  & \beta (M\!-\!TS_W)=\beta  \int^{(\mathrm{BH})}_{S_\infty} (\boldsymbol{Q}_{\xi_t}\!+\!\xi_t \cdot \boldsymbol{B} )\!-\!  \beta\int^{(\mathrm{BH})}_{S_h} \boldsymbol{Q}_{\xi_t} \\ 
  & \hspace{2cm}   -\beta  \int^{(\mathrm{bg})}_{S_\infty} (\boldsymbol{Q}_{\xi_t}+\xi_t \cdot \boldsymbol{B}  ) \\
  & \hspace{1.2cm}  =\beta \left(-\int^{(\mathrm{BH})}_{ \Sigma } \xi_t \cdot \boldsymbol{L}+\int^{(\mathrm{BH})}_{S_\infty}\xi_t\cdot \boldsymbol{B}\right)\\ 
  &\hspace{2cm} -\beta\left(- \int^{(\mathrm{bg})}_{ \Sigma } \xi_t \cdot \boldsymbol{L}+ \int^{(\mathrm{bg})}_{S_\infty}\xi_t\cdot \boldsymbol{B}\right)\\
 &\hspace{1.2cm}=-\beta \int^{(\mathrm{Reg.})}_{ \Sigma } \xi_t \cdot \boldsymbol{L}+\beta \int^{(\mathrm{Reg.})}_{S_\infty}\xi_t\cdot \boldsymbol{B}\\
  & \hspace{1.2cm}=  \int^{(\mathrm{Reg.})}_{ \mathcal{M} }\boldsymbol{L}_E+\int^{(\mathrm{Reg.})}_{\partial \mathcal{M}} \boldsymbol{B}_E \equiv I_E. \label{sec2qsr}
\end{split}
\end{align}
In the first step, we use the definition of $M$ in Eq.~\eqref{sec2mass} and the key result from the Iyer--Wald formalism: $\int^{(\mathrm{BH})}_{S_h} \boldsymbol{Q} = T S_W$, where $S_W$ denotes the well-known Wald entropy \cite{Iyer:1994ys,Iyer:1995kg}. The second step invokes the integral form of Eq.~\eqref{ff2}, which means $\oint_S \boldsymbol{Q} = -\int_\Sigma \xi_t \cdot \boldsymbol{L}$. For the third step, we simply abbreviate the expressions using the regularized integral $\int^{(\mathrm{Reg.})} \equiv \int^{(\mathrm{BH})} - \int^{(\mathrm{bg})}$. In the final step, we perform a Wick rotation $t = -i\tau$, with $\boldsymbol{L}_E = -i\boldsymbol{L}$ and $\boldsymbol{B}_E = -i\boldsymbol{B}$. Specifically, we have
$-\beta \int \xi_t \cdot \boldsymbol{L} = \int d\tau \wedge \frac{\partial}{\partial \tau} \cdot \boldsymbol{L}_E = \int \boldsymbol{L}_E$. A similar derivation applies to $\boldsymbol{B}_E$, with an additional sign flip caused by the orientation convention of the boundary $\Gamma$ (denoted as $\partial \mathcal{M}$ as well).

This derivation confirms the validity of the Euclidean action method. We formally define the Euclidean action as:
\begin{align}
   I_E \equiv  \int^{(\mathrm{Reg.})}_{ \mathcal{M} }\boldsymbol{L}_E+\int^{(\mathrm{Reg.})}_{\partial \mathcal{M}} \boldsymbol{B}_E. \label{sec2eucl}
\end{align}
Based on our preceding analysis, Eq.~\eqref{sec2qsr} establishes the quantum statistical relation 
\begin{align}
    I_E = \beta M - S_W,
\end{align}
while Eq.~\eqref{sec2firstlaw} enforces the first law of thermodynamics
\begin{align}
    \delta M = T \delta S.\label{dmts}
\end{align}
Combining these two results guarantees that thermodynamic quantities can be extracted directly from $I_E$. For instance, the mass is obtained as: $\frac{\partial I_E}{\partial \beta} = M - \beta \frac{\partial M}{\partial \beta} + \frac{\partial S_W}{\partial \beta} = M$ (the last two terms cancel due to Eq.\eqref{dmts}). Note that we implicitly identify \(S_W\) as the black hole entropy $S$, except for the specific scenario discussed in Sec.\ref{subsec3}.

 For asymptotically AdS black holes in Einstein gravity and higher-derivative gravity theories, Ref.~\cite{Guo:2025ohn} shows that the boundary term always vanishes after subtraction, following a careful examination of the asymptotic behaviors of such terms. This means that when using Eqs.~\eqref{sec2mass} and~\eqref{sec2eucl} to compute the mass and Euclidean action, one may safely ignore $\boldsymbol{B}_E$. We remark that this property holds for pure gravity in AdS spacetime. Once we go beyond this regime, such as in matter-coupled theories, we should carefully re-examine possible boundary contributions.

\section{Applying the subtraction method to matter–coupled gravity}\label{sec:III}

In last section, we have shown that for pure gravity, the background subtraction method directly follows from the Iyer–Wald formalism and yields correct thermodynamics. The reasoning is quite general, so there is no obvious reason why this conclusion should fail when matter fields are present. For a diffeomorphism-invariant Lagrangian, inclusion of matter only modifies the explicit expressions of $\boldsymbol{Q}$ and $\boldsymbol{\Theta}$ by adding matter-related contributions, and none of the general steps in Sec. \ref{sec20} are altered.

In fact, the key step in transitioning from the Iyer–Wald formalism to the Euclidean background subtraction method is rewriting Eq.~\eqref{iwadsbh} into the form of Eq.~\eqref{sec2firstlaw}, which are essentially equivalent expressions. It is therefore natural to expect that the Iyer–Wald formalism and the Euclidean subtraction method are equally effective in analyzing black hole thermodynamics. Whenever the Iyer–Wald formalism applies cleanly and correctly reproduces the physical mass of the system, the Euclidean approach should also work smoothly. In Secs.~\ref{subsec1} and \ref{subsec2}, we make this expectation explicit through two representative examples.

On the other hand, in the scenarios where the Iyer–Wald formalism encounters difficulties, the Euclidean method cannot be expected to perform better. We discuss two relevant subtleties in Sec.~\ref{subsec3}.

\subsection{Higher-derivative gravity with non-minimally coupled Maxwell field}\label{subsec1}
We first consider an example taken from Sec.4.5.1 of Ref.~\cite{Feng:2015sbw}. This work analyzed the black hole thermodynamic behavior directly using the Iyer–Wald formalism and found inconsistencies with the results obtained via the Euclidean action in Ref.~\cite{Cai:2011uh}. This led the authors to assert that their example serves as a lesson that the quantum statistical relation of the Euclidean action approach becomes problematic in theories with non-minimally coupled matter fields.

Naturally, we aim to use our diagnostic framework to identify where the Euclidean action approach might fail. Interestingly, we find that the Euclidean method works perfectly in this example and no issues arise. We present the diagnostic process below.

We start with the following Lagrangian:
\begin{align}
\boldsymbol{L} = \frac{1}{16\pi}\big(R - 2\Lambda - \frac{1}{4}F^2 + \gamma \mathcal{L}_{\text{hd}}\big)\boldsymbol{\epsilon}, \label{sec31lag}
\end{align}
where $\mathcal{L}_{\text{hd}}$ includes the non-minimal coupling terms between gravity and the Maxwell field:
\begin{align}
\mathcal{L}_{\text{hd}} = RF^2 - 4R_{\mu\nu}F^{\mu\rho}F^{\nu}_{\ \rho} + R_{\mu\nu\rho\sigma}F^{\mu\nu}F^{\rho\sigma}.
\end{align}

The black hole solution can be solved with the form
\begin{align}
ds^2 = -h(r)dt^2 + \frac{dr^2}{f(r)} + r^2d\Omega_{2,k}, \label{sec31ds}
\end{align}
where $d\Omega_{2,k} = \frac{dx^2}{1 - kx^2} + (1 - kx^2)dy^2$. Here, $k = 1, 0, -1$ corresponds to black holes with spherical, planar, or hyperbolic horizons, respectively. For non-compact cases ($k = 0, -1$), $S_r$ should be interpreted as a codimension-2 surface spanned by $x$ and $y$, and the physical quantities discussed below should be understood as densities per unit area (normalized to $4\pi$).

Treating the $\gamma \mathcal{L}_{\text{hd}}$ term as a perturbation, to first order in $\gamma$, the black hole solution can be obtained around the original Reissner–Nordström (RN) metric and is given by
\begin{align}
f(r) = f^{(0)}(r)\left(1 + \gamma f^{(1)}(r)\right), \quad h(r) = \left(1 + \frac{\gamma q^2}{r^4}\right)f(r), \label{sec21bhsol}
\end{align}
where
\begin{align}
h^{(0)}(r) & = f^{(0)}(r) = k + \frac{q^2}{4r^2} - \frac{\Lambda r^2}{3} - \frac{\bar{\mu}}{r},\\
f^{(1)}(r) &  \!= \! -  \frac{q^2}{2r^4}  \!+ \! \frac{1}{f^{(0)}(r)}\Bigg( \frac{c_1}{4r}  \!+ \! \frac{q^2(20kr^2  \!+ \! q^2)}{40r^6} \nonumber \\
&\quad + \frac{\Lambda q^2}{2r^2} \Bigg).
\end{align}
The electromagnetic field is described by $A_\mu = \{\phi(r), 0, 0, 0\}$, with:
\begin{align}
\phi(r) = -\frac{q}{r} + \frac{\gamma q\left(9q^2 - 80\Lambda r^4 - 60\bar{\mu}r\right)}{30r^5}.
\end{align}
The mass parameter $\bar{\mu}$ and constant $c_1$ are related to the horizon radius $r_h$ and charge $q$ by the requirement $f(r_h) = 0$. Their expressions are: $\bar{\mu} = kr_h + \frac{q^2}{4r_h} - \frac{\Lambda r_h^3}{3}$, and
$c_1 = -\frac{2kq^2}{r_h^3} - \frac{q^4}{10r_h^5} - \frac{2\Lambda q^2}{r_h}$.

Starting from the Lagrangian, we derive the Iyer–Wald surface charge density for this system as:
\begin{align}
\boldsymbol{k}_{\xi} = \delta \mathbf{Q}_\xi - \xi \cdot \boldsymbol{\Theta}[\delta \phi],
\end{align}
where
\begin{align}
 &  \  \boldsymbol{Q}_\xi= \boldsymbol{Q}_\xi^{(g)} + \boldsymbol{Q}_\xi^{(em)} \nonumber \\
&= \! \left(  \!- \! 2P^{\mu\nu\rho\sigma}\nabla_\rho \xi_\sigma  \!+ \! 4\xi_\sigma \nabla_\rho P^{\mu\nu\rho\sigma}  \!+ \! 4\mathcal{G}^{\mu\nu}\xi^\rho A_\rho \right)\boldsymbol{\epsilon}_{\mu\nu..},
\end{align}
and
\begin{align}
\boldsymbol{\Theta} &= \left( 2P^{\mu\alpha\beta\nu}\nabla_\nu \delta g_{\alpha\beta} - 2\nabla_\nu P^{\mu\alpha\beta\nu}\delta g_{\alpha\beta} \right. \nonumber \\
&\quad \left. + 4\mathcal{G}^{\mu\nu}\delta A_{\nu} \right)\boldsymbol{\epsilon}_{\mu\cdots}.
\end{align}
Here, the symbols $P^{\mu\nu\rho\sigma}$ and $\mathcal{G}^{\mu\nu}$ are defined as
\begin{align}
\begin{split}
&  P^{\mu\nu\rho\sigma} \equiv \frac{\partial \mathcal{L}}{\partial R_{\mu\nu\rho\sigma}} =\frac{1}{16\pi}\big( g^{\mu[\rho}g^{\sigma]\nu}  \\&  +  \gamma ( g^{\mu[\rho}g^{\sigma]\nu}F^2 + g^{\rho[\mu}F^{\nu]\alpha}F^{\sigma}_{\ \alpha} + F^{\mu\nu}F^{\rho\sigma} ) \big),
\end{split}
\end{align}
\begin{align}
\begin{split}
& \mathcal{G}^{\mu\nu} \equiv \frac{\partial \mathcal{L}}{\partial F_{\mu\nu}}
= \frac{1}{16\pi}\big( -\frac{1}{4}F^{\mu\nu}  \nonumber \\ &+ \gamma \left( RF^{\mu\nu} - 4R_\rho^{\ \mu}F^{\rho\nu} + R^{\mu\nu\rho\sigma}F_{\rho\sigma} \right) \big).
\end{split}
\end{align}
At the boundary $\Gamma$, we decompose $\boldsymbol{\Theta}\big|_{\Gamma}$ following the standard procedure \cite{Padmanabhan:2014lwa,Chakraborty:2017zep,Jiang:2018sqj}:
\begin{align}
\boldsymbol{\Theta}\big|_{\Gamma} = -\delta \boldsymbol{B} + d\boldsymbol{C}[\delta\phi] + \boldsymbol{F}[\delta\phi].
\end{align}
This decomposition is tedious but straightforward, leading to the explicit forms of $\boldsymbol{B}$ and $\boldsymbol{F}[\delta\phi]$:
\begin{align}
& \boldsymbol{B} = 4\mathcal{P}_{\alpha\beta}K^{\alpha\beta}\hat{\boldsymbol{\epsilon}}, \label{sec31boun} \\
& \boldsymbol{F}[\delta\phi] = \bigg( 4K^{\alpha\beta}\delta\mathcal{P}_{\alpha\beta} + \big( 2n^{\nu}\nabla^{\mu}P_{\alpha\mu\nu\beta} + 6\mathcal{P}_{\mu\alpha}K^{\mu}_{\ \beta} \nonumber \\
&\quad - 2\mathcal{P}_{\mu\nu}K^{\mu\nu}h_{\alpha\beta} \big)\delta h^{\alpha\beta}  + 4n_{\mu}\mathcal{G}^{\mu\nu}\delta A_{\nu} \Bigg)\hat{\boldsymbol{\epsilon}},
\end{align}
where $\mathcal{P}^{\mu\nu}$ is defined as
\begin{align}
\mathcal{P}^{\mu\nu} \equiv P^{\alpha\beta\gamma\delta}n_{\beta}n_{\delta}h_{\alpha}^{\ \mu}h_{\gamma}^{\ \nu}.
\end{align}

We now take the timelike Killing vector $\xi_t = \frac{\partial}{\partial t}$. The identity $\int_{S_h}\boldsymbol{k}_{\xi_t} = \int_{S_\infty}\boldsymbol{k}_{\xi_t}$ produces
\begin{align}
T\delta S + \Phi\delta Q = \int_{S_\infty}^{(\mathrm{BH})} \delta\left( \boldsymbol{Q}_{\xi_t} + \xi_t \cdot \boldsymbol{B} \right) - \xi_t \cdot \boldsymbol{F}[\delta\phi]. \label{sec31firstlaw}
\end{align}
Next, we introduce the reference background spacetime. For the metric in Eq.~\eqref{sec21bhsol}, we set $r_h = q = 0$ and apply a redshift factor to ensure the black hole spacetime and background behave consistently as $r=r_c \to \infty$. For this horizonless background, the identity $\int_{S_\infty}\boldsymbol{k}_{\xi_t} = 0$ holds, leading to:
\begin{align}
0 = \int_{S_\infty}^{(\mathrm{bg})} \delta\left( \boldsymbol{Q}_{\xi_t} + \xi_t \cdot \boldsymbol{B} \right) - \xi_t \cdot \boldsymbol{F}[\delta\phi]. \label{sec31firstlaw2}
\end{align}
Subtracting Eq.~\eqref{sec31firstlaw2} from Eq.~\eqref{sec31firstlaw}, we find that the $\boldsymbol{F}$-term contributions cancel in the limit $r_c \to \infty$:
\begin{align}
\int_{S_{r_c}}^{(\mathrm{Reg.})}\xi_t \cdot \boldsymbol{F}[\delta\phi]\sim \mathcal{O}\left(\frac{1}{r_c^2}\right) \to 0.
\end{align}
This leads to
\begin{align}
T\delta S + \Phi\delta Q = \delta \int_{S_\infty}^{(\mathrm{Reg.})} \left( \boldsymbol{Q}_{\xi_t} + \xi_t \cdot \boldsymbol{B} \right). \label{sec31firstReg}
\end{align}
By comparing Eq.~\eqref{sec31firstReg} with the standard first law of black hole thermodynamics $T\delta S + \Phi\delta Q = \delta M$, we recognize the physical mass as
\begin{align}
M \equiv \int_{S_\infty}^{(\mathrm{Reg.})} \left( \boldsymbol{Q}_{\xi_t} + \xi_t \cdot \boldsymbol{B} \right).
\end{align}
Crucially, Eq.~\eqref{sec31firstReg} is simply a formally equivalent rewrite of Eq.~\eqref{sec31firstlaw}. Thus, the $\delta M$ extracted here is guaranteed to be consistent with the Iyer–Wald formalism.

In addition, following the derivation in Sec.\ref{sec20}, we immediately obtain the quantum statistical relation for the Euclidean action:
\begin{align}
I_E \equiv \int_{\mathcal{M}}^{(\mathrm{Reg.})}\boldsymbol{L}_E + \int_{\partial\mathcal{M}}^{(\mathrm{Reg.})}\boldsymbol{B}_E = \beta\left( M - T S_W - \Phi Q \right). \label{sec31qsr}
\end{align}
Combining Eq.~\eqref{sec31qsr} with Eq.~\eqref{sec31firstReg} allows us to extract thermodynamic quantities (mass, entropy, charge) directly from $I_E$. Therefore, our diagnostic analysis confirms that the Euclidean action approach operates smoothly in this scenario, with no ambiguities or failures. Besides, our conclusion does not depend on the horizon topology ($k = 1, 0, -1$), as our analysis imposes no restrictions on $k$.

To further validate the Euclidean method, we explicitly compute the Euclidean action and demonstrate that it reproduces the same thermodynamic quantities as the Iyer–-Wald formalism. The Euclidean action is defined as:
\begin{align}
I_E \equiv \int_{\mathcal{M}}^{(\mathrm{Reg.})}\boldsymbol{L}_E + \int_{\partial\mathcal{M}}^{(\mathrm{Reg.})}\boldsymbol{B}_E,
\end{align}
where $\boldsymbol{L}$ and $\boldsymbol{B}$ are given by Eqs.~\eqref{sec31lag} and \eqref{sec31boun}, respectively. As noted in the Introduction, the computation simplifies to inserting the black hole solution into the Lagrangian and evaluating the integral. To first order in $\gamma$, the free energy $G = I_E/\beta$ is obtained as
\begin{align}
G = \frac{12k r_h^2 \!-\! 3q^2 \!+\! 4\Lambda r_h^4}{48r_h} \!-\! \frac{\gamma q^2\left(q^2\!-\!20k r_h^2  \!+\! 20\Lambda r_h^4\right)}{160r_h^5}.
\end{align}

 For the metric in Eq.~\eqref{sec31ds}, the temperature is derived from the surface gravity at $r = r_h$, yielding:
\begin{align}
T &= \frac{\sqrt{h'(r_h)f'(r_h)}}{4\pi} \nonumber \\
&= \frac{4k r_h^2 \!-\! q^2 \!-\! 4\Lambda r_h^4}{16\pi r_h^3} - \frac{\gamma q^2\left(12k r_h^2 \!+\! q^2 \!+\! 4\Lambda r_h^4\right)}{32\pi r_h^7}.
\end{align}

The electric potential at the horizon is calculated from $\Phi = -\xi_t^\mu A_\mu$, resulting in:
\begin{align}
\Phi = -\frac{q}{r_h} + \frac{\gamma q\left(9q^2 - 80\Lambda r_h^4 - 60\bar{\mu}r_h\right)}{30r_h^5}.
\end{align}

Using the thermodynamic relation for free energy: $dG = -SdT - Qd\Phi$, we extract the charge $Q$ and entropy $S$ by taking partial derivatives:
\begin{align}
Q &= -\frac{\partial G}{\partial \Phi} = \frac{q}{4}, \\
S &= -\frac{\partial G}{\partial T} = \pi r_h^2.
\end{align}
The charge $Q$ matches the conserved charge computed directly from  $Q = -\frac{1}{4\pi}\int_S \mathcal{G}^{\mu\nu}\boldsymbol{\epsilon}_{\mu\nu..}$, and the entropy $S$ agrees with that from Wald entropy formula. Finally, the mass $M$ is recovered from the thermodynamic relation $M = G + TS + Q\Phi$, which simplifies to
\begin{align}
M = \frac{\bar{\mu}}{2} - \frac{\gamma}{8}c_1.
\end{align}
All the above results from the Euclidean background subtraction method are identical to those obtained via the Iyer–Wald formalism in Ref.~\cite{Feng:2015sbw}. This again verifies that the Euclidean method yields fully consistent thermodynamic quantities even for gravity theories with non-minimally coupled matter fields. 

\subsection{Gravity coupled with a background Kalb-Ramond field}\label{subsec2}

For pure gravity theories in AdS spacetime, Ref.~\cite{Guo:2025ohn} shows that boundary terms vanish after background subtraction. Consequently, the Euclidean action is often computed using the simplified formula
\begin{align}
I_E \equiv \int_{\mathcal{M}}^{(\mathrm{Reg.})}\boldsymbol{L}_E. \label{sec2i_el}
\end{align}
This simplification also holds in many matter-coupled cases, for example, the Maxwell field example in Sec.~\ref{subsec1} yields $\int_{\partial\mathcal{M}}^{(\mathrm{Reg.})}\boldsymbol{B}_E = 0$.

However, we now present a case that serves as a cautionary note: when moving beyond the regime of pure gravity or AdS spacetimes, Eq.~\eqref{sec2i_el} may fail to yield correct thermodynamic quantities. This occurs when matter fields possess non-vanishing vacuum expectation values that approach finite constants at infinity, leading to nonzero boundary term contributions. We illustrate this with gravity coupled to a Kalb-Ramond field, which is a second-rank antisymmetric tensor field from the bosonic spectrum of string theory. As far as we know, Euclidean action computations for such black holes have not been reported previously.

The Lagrangian for Einstein gravity coupled to a Kalb-Ramond field is $\boldsymbol{L} = L\boldsymbol{\epsilon}$ \cite{Altschul:2009ae,Liu:2025fxj}, where:
\begin{align}
L &= \frac{1}{16\pi}\big( R - 2\Lambda + \gamma B^{\rho\mu}B^{\nu}_{\ \mu}R_{\rho\nu} \big) \nonumber \\
&\quad - \frac{1}{12}H^{\mu\nu\rho}H_{\mu\nu\rho} - V\left(X\right). \label{krfield}
\end{align}
Here, $B_{\mu\nu}$ is the Kalb-Ramond field, $H_{\mu\nu\rho} \equiv \nabla_{[\mu}B_{\nu\rho]}$ is its field strength, and $V(X)$ is a potential with $X = B^{\mu\nu}B^{\mu\nu} \pm b^2$. The constant $b^2$ controls the vacuum expectation value of $B_{\mu\nu}$ through the condition $V(X) = 0$.

We study the 4-dimensional black hole solution from Ref.~\cite{Liu:2025fxj}, which is obtained for the potential $V(X) = \frac{\lambda}{2} X$ with $\lambda = \frac{\Lambda \gamma}{8\pi\left(1 - \frac{\gamma b^2}{2}\right)}$. The metric takes the form:
\begin{align}
ds^2 = -f(r)dt^2 + \frac{dr^2}{g(r)} + r^2d\Omega_2,
\end{align}
where $f(r) = g(r)$, and
\begin{align}
f(r) = \frac{1}{1 - \frac{\gamma b^2}{2}} - \frac{2m}{r} - \frac{\Lambda r^2}{3\left(1 - \frac{\gamma b^2}{2}\right)}.
\end{align}
The non-zero components of the Kalb-Ramond field's vacuum expectation value are $\langle B_{10} \rangle = -\langle B_{01} \rangle = \frac{b}{\sqrt{2}}$. The corresponding field strength $H_{\mu\nu\rho}$ vanishes, so we can omit its effects here. The mass parameter $m$ is related to the horizon radius $r_h$ by $m = \frac{r_h\left(3 - \Lambda r_h^2\right)}{6 - 3\gamma b^2}$, which is enforced by the condition $f(r_h) = 0$.

Next we compute the Euclidean action:
\begin{align}
I_E = \int_{\mathcal{M}}^{(\mathrm{Reg.})}\boldsymbol{L}_E + \int_{\partial\mathcal{M}}^{(\mathrm{Reg.})}\boldsymbol{B}_E. \label{sec32lag}
\end{align}
A consistent variational principle requires the boundary term $\boldsymbol{B}$ to take the form:
\begin{align}
\boldsymbol{B} = \frac{1}{8\pi}\big( K + \frac{1}{2}\gamma B^{\mu\rho}B^{\nu}_{\ \rho}\left(K_{\mu\nu} + n_{\mu}n_{\nu}K\right) \big)\hat{\boldsymbol{\epsilon}},
\end{align}
As usual, $K$ denotes the extrinsic curvature of the boundary, and $n_{\mu}$ is the unit normal vector.

Substituting the black hole solution into Eq.~\eqref{sec32lag}, we evaluate the two integrals separately. First, the regularized bulk integral,
\begin{align}
\int_{\mathcal{M}}^{(\mathrm{Reg.})}\boldsymbol{L}_E = \frac{\beta r_h\left(-3\gamma b^2 + \Lambda r_h^2 + 3\right)}{6\left(2 - \gamma b^2\right)}.
\end{align}
Second, the regularized boundary integral,
\begin{align}
\int_{\partial\mathcal{M}}^{(\mathrm{Reg.})}\boldsymbol{B}_E = \frac{\beta b^2 r_h \gamma\left(-\Lambda r_h^2 + 3\right)}{12\left(2 - \gamma b^2\right)}.
\end{align}
Clearly, the boundary term contributes a non-vanishing finite piece. Summing them together gives the total Euclidean action
\begin{align}
I_E = \frac{1}{12}\beta r_h\left(\Lambda r_h^2 + 3\right).
\end{align}
The free energy is obtained as $F = I_E/\beta$. To confirm consistency, we compute the Hawking temperature from the horizon surface gravity:
\begin{align}
T = \frac{\sqrt{f'(r_h)g'(r_h)}}{4\pi} = \frac{1 - \Lambda r_h^2}{4\pi r_h - 2\pi \gamma b^2 r_h}.
\end{align}
Then, using the thermodynamic relations, we extract the entropy and mass as
\begin{align}
S &= -\frac{\partial F}{\partial T} = \pi r_h^2\left(1 - \frac{\gamma b^2}{2}\right), \\
M &= F + TS = m\left(1 - \frac{\gamma b^2}{2}\right).
\end{align}
The results match those from the direct application of the Iyer–Wald formalism, confirming that the full Euclidean action (including boundary terms) is necessary for consistency in this case.

\subsection{Two subtleties in applying the Iyer–Wald formalism and background subtraction method}\label{subsec3}
As emphasized before, the background subtraction method and Iyer–Wald formalism are equally effective for analyzing black hole thermodynamics. When the Iyer–Wald formalism encounters difficulties (a rare occurrence), the Euclidean method faces the same challenges. Below, we discuss two key subtleties.

\subsubsection{Non-integrable variations from matter field parameters}

In matter-coupled gravity, variations of matter field parameters contribute to $\boldsymbol{k}_{\xi_t} = \delta\mathbf{Q}_{\xi_t} - \xi_t \cdot \boldsymbol{\Theta}[\delta\phi]$, which can lead to non-integrable variations, i.e., expressions that cannot be written as the differential of a single quantity. 

For gauge fields, this issue is often resolved by a gauge choice: for example, if the gauge field $A_\mu$ is required to vanish at the horizon, one finds $\int_{\infty}\boldsymbol{k}_{\xi_t} = \delta M - \Phi\delta Q$, which is a non-integrable variation. But redefining the gauge can shift the $\Phi\delta Q$ term from the asymptotic boundary to the horizon, neatly yielding $\int_{\infty}\boldsymbol{k}_{\xi_t} = \delta M$.

However, for systems lacking such gauge symmetry, this resolution may not apply; whether a similar manipulation exists depends on the specific theory. A notable example is the thermodynamics of AdS black holes with scalar hair studied in Ref.~\cite{Li:2020spf}, where the author found
\begin{align}
\int_{\infty}\boldsymbol{k}_{\xi_t} \sim -2\delta f_v + \phi_s \delta\phi_v + 2\phi_v \delta\phi_s, \label{sec33w}
\end{align}
where $f_v$ is the metric's mass parameter, and $\phi_v$, $\phi_s$ characterize the scalar field (see Sec.5 of Ref.~\cite{Li:2020spf}). The term $\phi_s \delta\phi_v + 2\phi_v \delta\phi_s$ is non-integrable. Since a non-integrable expression and an integrable expression are fundamentally distinct, no choice of background can rewrite Eq.\eqref{sec33w} as an integrable variation. Indeed, for this example, our regularized calculations show that
\begin{align}
& \int_{\infty}^{(\mathrm{Reg.})}\delta (\boldsymbol{Q}+\xi_t \cdot \boldsymbol{B} )\sim \delta (-2f_v),\\
& \int_{\infty}^{(\mathrm{Reg.})}\xi_t \cdot \boldsymbol{F}[\delta\phi] \sim \phi_s \delta\phi_v + 2\phi_v \delta \phi_s \neq0.
\end{align}

This non-integrability does not indicate a failure of the Iyer--Wald formalism and the subtraction method. Actually, they still yield the correct thermodynamic relations among variations of different quantities involving $\delta f_v$, $\delta \phi_v$, and $\delta \phi_s$. The main problem is that ambiguities arise when identifying $\delta M$ from $\int_{\infty}\boldsymbol{k}_{\xi_t}$: neither the Iyer–Wald formalism nor the background subtraction method provides a clear criterion for classifying these terms. Accordingly, one must resort to an alternative approach. In this regard, Ref.~\cite{Li:2020spf} resolves the issue using holographic renormalization, properly rewriting the integral \eqref{sec33w} as
\begin{align}
\int_{\infty}\boldsymbol{k}_{\xi_t} \sim \delta\mathcal{E} + \langle O \rangle d\phi_s,
\end{align}
where $\mathcal{E}$ is the renormalized energy extracted from the holographic stress tensor, and $\langle O \rangle$ is interpreted as the response to the source $\phi_s$.

\subsubsection{Discrepancy between black hole entropy and Wald entropy}
The black hole entropy $S$ is defined via $\int_{S_h}\boldsymbol{k}_{\xi} = T\delta S$, and the Wald entropy $S_W$ is defined via $\int_{S_h}\mathbf{Q}_\xi = TS_W$, where $\xi$ has to be taken as the horizon Killing vector of the spacetime under study. These two quantities coincide with each other, differing only in extreme cases where matter fields diverge at the horizon \cite{An:2024fzf,Feng:2015oea}.

Concretely, the equality between $S$ and $ S_W$ relies on the horizon Killing vector vanishing at the bifurcation surface, which eliminates the $-\xi \cdot \boldsymbol{\Theta}[\delta\phi]$ term in $\boldsymbol{k}_{\xi}$. However, if a matter field diverges at $r = r_h$, it can counteract this effect and yield a non-trivial contribution to $\int_{S_h}\boldsymbol{k}_{\xi}$, causing the inequality $S \neq S_W$. When this situation arises, it brings about difficulties in consistently interpreting black hole thermodynamics. The awkward problem of $S_W \neq S$ (within the Iyer–Wald formalism) translates to $I_E = \beta (M - TS_W) \neq \beta (M - TS)$ (within the Euclidean action method). In such cases, $I_E$ cannot be legitimately related to the system's free energy, and we thus cannot extract the various thermodynamic quantities of the black hole from $I_E$.

To our knowledge, there is no general solution to this issue. A tentative proposal from Ref.~\cite{Wu:2025jes} suggests that black hole solutions extremizing the Euclidean action may possess conical singularities. Anyway, such cases appear to be rare and do not undermine the applicability of the general formalisms in  most practical scenarios.

\section{Concluding remarks} \label{sec:conclusion}
In this work, we revisit the background subtraction method for computing the Euclidean action of black holes and clarify its validity in gravity theories coupled to matter fields.

As demonstrated in Sec.~\ref{sec20}, the background subtraction method naturally follows from the Iyer--Wald formalism in pure gravity scenarios. When implemented rigorously, the obtained Euclidean action necessarily satisfies the quantum statistical relation, regardless of the detailed structure of the theory, and the Euclidean method is guaranteed to reproduce consistent thermodynamic results.  

We note here that, although we focus on AdS spacetimes in Sec.~\ref{sec20}, all of our analysis also applies to asymptotically-flat spacetimes. In such scenarios, the boundary term $\int_{\partial\mathcal{M}}^{(\mathrm{Reg.})}\boldsymbol{B}_E$ yields the dominant contribution to the Euclidean action $I_E$. For example, for a Schwarzschild black hole in Einstein gravity, the bulk action vanishes identically, and the Euclidean action takes the form $\int_{\partial\mathcal{M}}^{(\mathrm{Reg.})}\boldsymbol{B}_E \sim \int_{\partial M}(K-K_0)\sqrt{|h|}d^3x$. We emphasize that the last term involves $K_0\sqrt{|h|}$ rather than $K_0\sqrt{|h_0|}$, as the background spacetime and the black hole spacetime share the same asymptotic geometry at infinity, which is the key property guaranteed by the redshift factor.

In Sec.~\ref{sec:III}, we further extend this analysis to representative matter-coupled gravity models and verify that the core conclusion persists. For the analyzed examples, after proper background subtraction, the Euclidean action yields thermodynamic quantities fully consistent with those derived from the Iyer--Wald formalism. 

We also identify two subtleties requiring caution in practical applications: (i) non-integrable variations arising from matter field parameters, which can lead to ambiguities in identifying the physical mass; and (ii) discrepancies between black hole entropy and Wald entropy, occurring when matter fields diverge at the horizon and breaking the standard relation between the Euclidean action and free energy.

To summarize, the background subtraction method remains a reliable, efficient tool for black hole thermodynamics beyond pure gravity. Our results clarify when the Euclidean approach can be trusted in matter–coupled gravity theories and when additional care is required. We anticipate this work will provide a useful framework for future studies of black hole thermodynamics in higher-derivative gravities, holographic systems, and nontrivial matter-coupled scenarios.

\section*{Acknowledgments}
 YX would like to thank Hong-Bao Zhang, Xi-Yao Guo and Jie-Qiang Wu for the useful discussions. YX is also thankful to the Higgs Centre for Theoretical Physics at the University of Edinburgh for providing research facilities and hospitality during the visit. This work was supported in part by the National Natural Science Foundation of China with Grant No. 12475048, the Hebei Natural Science Foundation with Grant No. A2024201012,  the Science Research Project of Hebei Education Department with Grant No. JCZX2026019, and the China Scholarship Council with Grant No. 202408130101.


\end{CJK*}

\end{document}